\documentclass{emulateapj}

\slugcomment{\bf Accepted for publication in the Publications of the Astronomical Society of the Pacific}
\shorttitle{Optical Companion of HS~1136+6646}
\shortauthors{Liebert et al.}

\begin{document}

\title{The Pre-cataclysmic Binary HS~1136+6646 May Have a Companion}

\author{James Liebert\altaffilmark{1}, Kurtis A. Williams\altaffilmark{1,2}
J. B. Holberg\altaffilmark{3}, and D. K. Sing\altaffilmark{3} }

\altaffiltext{1}{Department of Astronomy and Steward Observatory,
University of Arizona, Tucson, AZ 85721, liebert@as.arizona.edu}

\altaffiltext{2}{Current address: Department of Astronomy, University
of Texas, Austin TX 78712, kurtis@astro.as.utexas.edu}

\altaffiltext{3}{Lunar and Planetary Laboratory, University of Arizona,
Tucson AZ 85721, holberg@vega.lpl.arizona.edu, singd@vega.lpl.arizona.edu}

\begin{abstract}

Because of the similarity of the primary star of HS~1136+6646 to the
planetary nebula central star BE Ursae Majoris, we did wide field
imaging of the former with an H$\alpha$ filter.  No nebulosity was
detected.  On the other hand, the point spread function of the star
appeared extended.  A partially-resolved red component is detected in
the image with the five-band Sloan Digital Sky Survey.  Most
importantly, a companion is easily resolved in the \emph{HST}
acquisition image for the published \emph{STIS} observation.  A
companion to the pre-cataclysmic binary is present at a separation of
1.349'' at position angle 54.4$^o$.  Evidence indicates that it is
likely of K spectral type.  We cannot demonstrate conclusively that this
component has common proper motion with the close binary.  However, the
similar apparent $z$ magnitudes and spectral types of HS~1136+6646B and
the resolved component make it likely that we have in reality a
hierarchial triple system.  In any case, the presence of this component
needs to be taken into account in future ground-based studies.

\end{abstract}

\keywords{white dwarfs -- binaries(including multiple): close --
  binaries: spectroscopic -- stars: individual(HS~1136+6646)}

\section{Introduction}

HS~1136+6646 (hereafter HS~1136) is a hot young DAO white dwarf plus K7V
secondary star, in a close detached binary.  The discovery from the
Hamburg Schmidt Survey was reported in Heber, Dreizler, \& Hagen (1996).
Sing et al. (2004) determined parameters for the individual components
and found an orbital period of 0.83607 days.  Radial velocity
measurements of both components were used to determine masses of
0.85$_\odot$ and 0.37M$_\odot$ for the primary and secondary,
respectively (Sing 2005).  The white dwarf parameters -- T$_{eff}$
$\sim$70,000~K and $log~g$ $\sim$7.75 -- could only be coarsely
estimated, due to the dependence of the parameters determined from
fitting the Balmer lines on the heavy element abundances.

Moreover, the optically-derived temperature was difficult to reconcile
with the far-UV spectrum of the Lyman line region. The \emph{Far Ultraviolet
Spectroscopic Explorer} spectrum shows the presence of O~VI
absorption lines and a spectral energy distribution whose slope persists
nearly to the Lyman limit, suggesting that the Balmer line T$_{eff}$
estimate is too low.  Fits to the Lyman lines by Good et al. (2004)
indicate a far higher value of 120,000~K -- among the highest
determinations for a DAO white dwarf -- with $log~g$ of 6.5.  These
parameters suggest that the system could have left the common envelope
even more recently than the Sing et al. (2004) estimate of
7.7$\times$10$^5$ years.  Indications are that the secondary star may be
overluminous compared to a main sequence star of its mass, still bloated
or out of thermal equilibrium from the common envelope phase.

The secondary is irradiated by strong UV radiation causing strong
variations in the strengths of the emission lines vs. orbital phase.  BE
Ursae Majoris (Ferguson et al. 1999, and references therein) is a
similar close DAO+K binary also showing a strong orbitally-modulated,
``reflection effect.''  Again, only coarse estimates of T$_{eff}$ and
$log~g$ were given by the above authors, suggesting that the primary
star could be similar to that of HS~1136.

An additional property of the BE~UMa system is the presence of a large
planetary nebula of low surface brightness (Liebert et al.1995).  As
part of a program by the second author of that paper to look for ancient
planetary nebulae (Tweedy \& Kwitter 1994), a 50 arc minute field around
BE~UMa was imaged in H$\alpha$ 6563\AA, [O~III] 5007\AA, and [N~II]
6584\AA.  The [N~II] observation yielded a nondetection, the [O~III] a
marginal one, but the H$\alpha$ image revealed the very faint nebula,
centered on the binary system, of about a 5 arc minute diameter.

This led us to try a similar observation of HS~1136, to see if the
systems were similar in yet another property.  In Section 2 we report
that no nebula was detected.  Instead, we made a different discovery,
discussed in Section 3.

\section {The Search for a Planetary Nebula}

\begin{figure}

\plotone{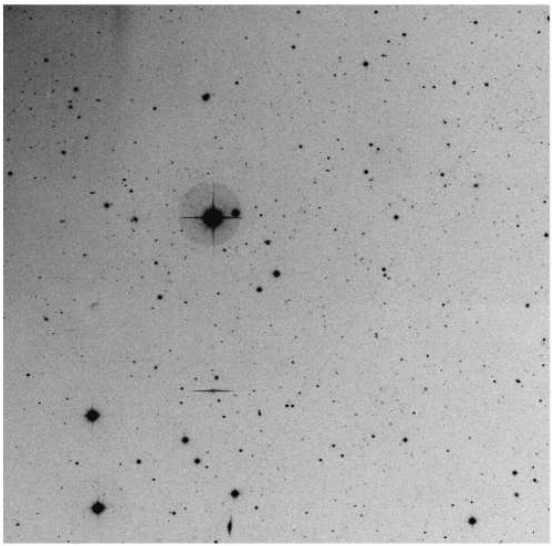}

\caption{The H$\alpha$ image centered on HS~1136 obtained with the
\emph{90Prime} camera.  Three 15-minute exposures were stacked to make
this combined image.  North is up, with east to the left.  The image
displayed is 15 arc minutes on a side. \label{fig1}}

\end{figure}

\emph{90Prime} is a wide field imager, mounted at the prime focus of the
Steward Observatory 2.3-m Bok reflector (Williams et
al. 2004)\footnote{We note that Ed Olszewski is the principal
investigator of the grant enabling construction}.  The detector is a
mosaic of four thinned Lockheed 4096$\times$4096 pixel CCDs.  The f/2.98
system provides a plate scale of 0.45$\arcsec$/pixel.  We centered the
binary star on Chip-1, the cosmetically best chip having no bad columns.
The resulting image through an H$\alpha$ filter is shown in Figure~\ref{fig1}.
Scattered light in the upper left of the figure is from a 6th magnitude
star, 3 Draconis, 25' northeast of HS~1136.  This complicates the
potential detection of faint nebulosity.  In any case, no evidence of
nebulosity is seen.  Because of the complication, we did not try to
estimate a sensitivity limit above which a nebula is not present.

One feature with this camera at prime focus is a continual need to
monitor the instrument focus.  When checking for this after the
observations with isophotal contour plots, it became apparent to us that
the target star appeared elongated at a position angle of 60 (or 240)
degrees from due north (Figure~2a), while several stars of similar
magnitude in the field appeared almostly perfectly circular (one of them
is shown in Figure~2b).  This strongly suggested to us that HS~1136 was
not a point source, but either would turn out to be a barely-resolved
source or a partially-resolved blend of two or more sources.  In
Figure~3 is shown the radial profiles for HS~1136 (solid points)
vs. those for four comparison stars.  That HS~1136 is more extended than
the others is indisputable.  The result was not due to the instrument
being out of focus.  However, the seeing at this time from extracting
the point spread functions was $\sim$1.6'' -- see Figure~3.

\begin{figure}

\plottwo{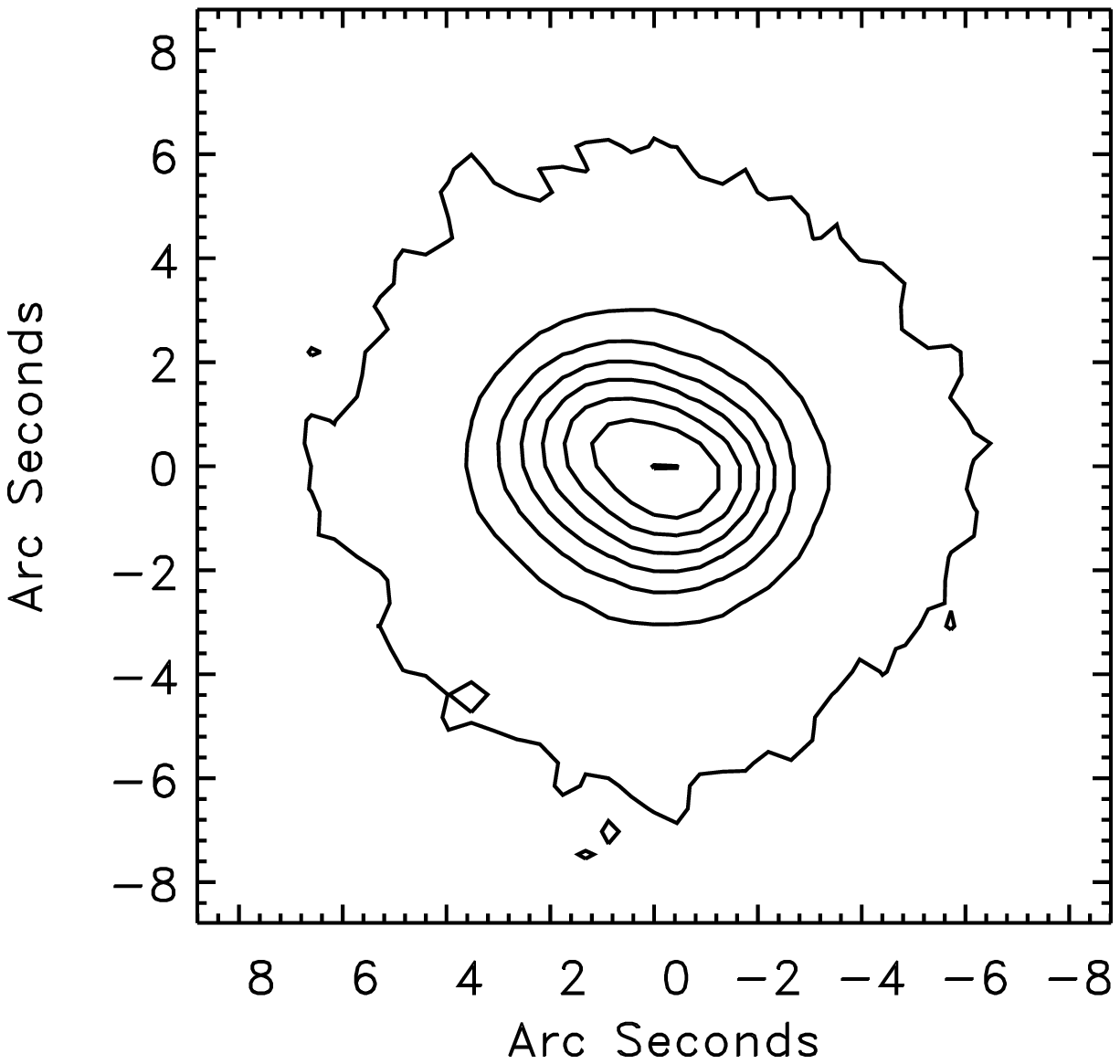}{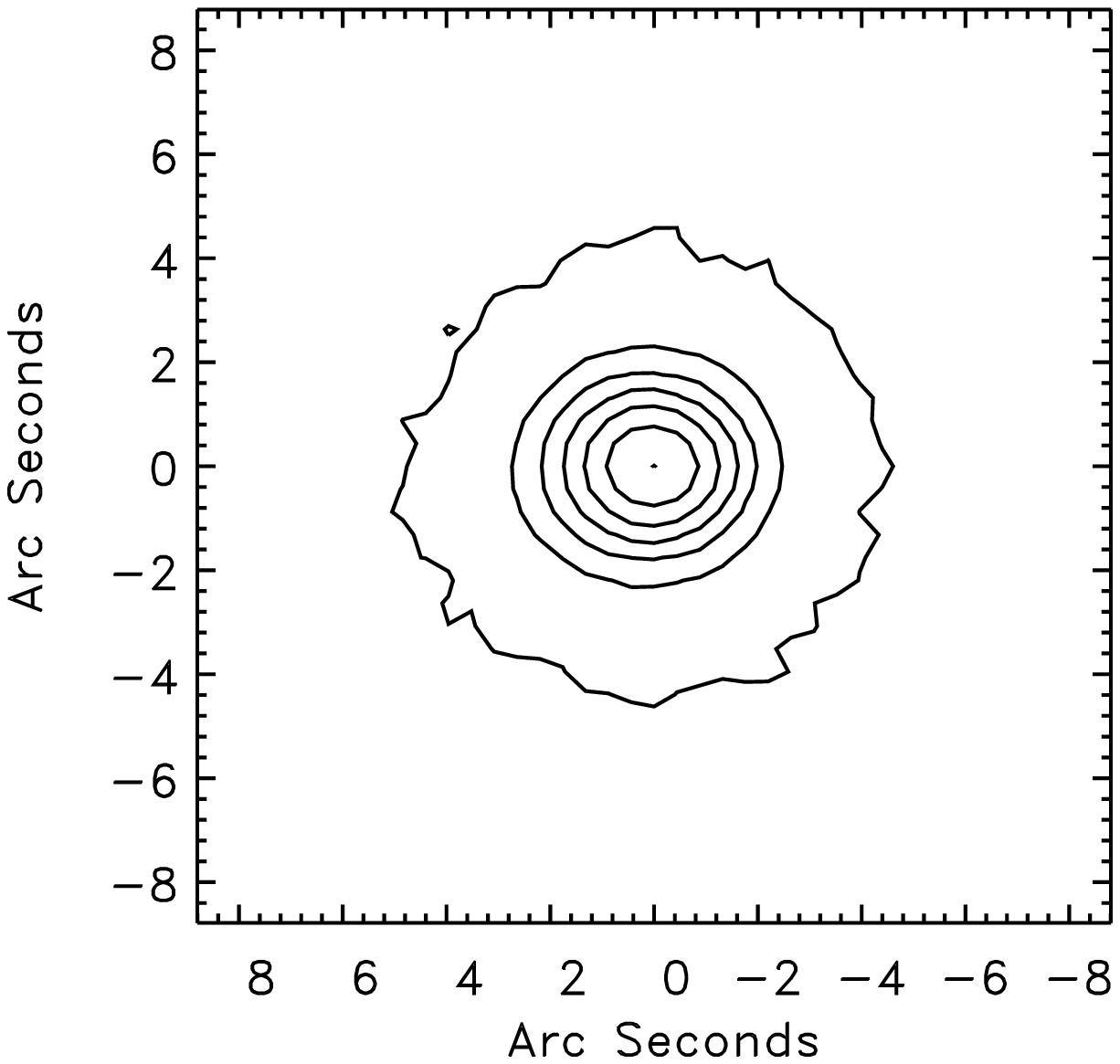}
\caption{The contour plots of counts, with each contour a successive
factor of two below the peak for (a, left) HS~1136 and (b, right) one
of the several reference stars in the field also measured.  North is
up and east is to the left.  That HS~1136 is extended relative to a
seeing-dominated point source is apparent.}

\end{figure}

\begin{figure}

\plotone{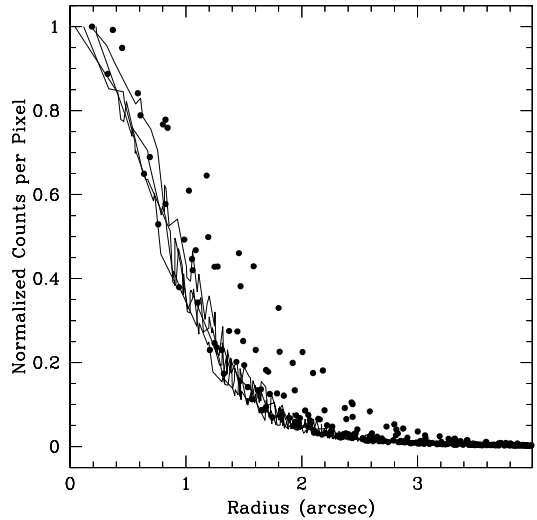}
\caption{The azimuthally-summed radial distribution of counts for
HS~1136 (filled circles) and four reference stars (dark curves).
The full width at half maximum of the latter indicates a seeing
of about 1.8 arc second.  The extension of the HS~1136 counts past
1'' is consistent with the presence of the close companion. }

\end{figure}

Had we discovered a very compact nebula in H$\alpha$ around this
unusual binary, instead of an old nebula?  A logical test was to
try a very short exposure in broad-band $V$, a wavelength range in
which only weak nebular emission could be anticipated.  We did
this observation (not shown here), but the HS~1136 PSF again showed
extension, the comparison stars did not.  Some other explanation
besides nebular emission had to be sought.

\section {The Result of Higher Resolution Imaging: A Companion}

The field is covered by the Sloan Digital Sky Survey (Gunn et al.  1998,
York et al. 2000, Stoughton et al 2002).  If one downloads the color
image from the DR5 Finding Chart Tool (Figure~4), the presence of a
fainter companion to HS~1136 at the same orientation as indicated by
Fig.~2a is apparent.  This image shows that the companion is much redder
than the optical light of the close binary, the latter dominated by the
very hot primary.  As we will show later, visual inspection of other
SDSS stellar sources suggests strongly that the companion is of K or M
spectral type.

\begin{figure}

\plotone{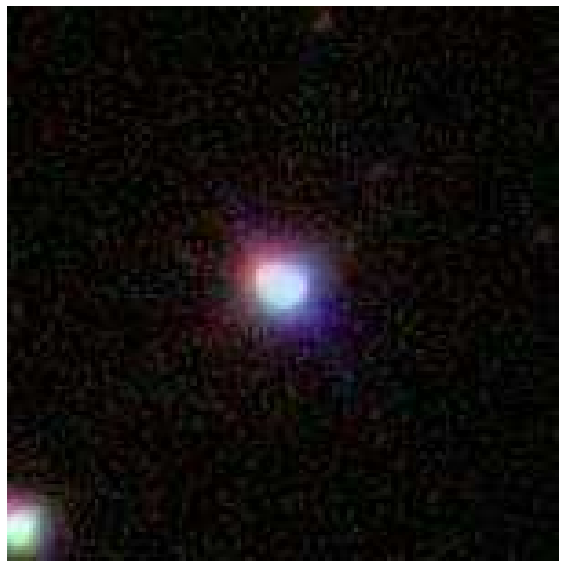}
\caption{The multi-color image of HS~1136 from the Sloan Digital
Sky Survey.  A faint red companion is marginally resolved 
from the blue-colored close binary.}

\end{figure}

The acquisition image taken for the previously-published spectrum with
the \emph{Space Telescope Imaging Spectrograph (STIS)} instrument on the
\emph{Hubble Space Telescope} is shown in Figure~5.  The USNO-B1 (Monet
et al. 2003) position of HS~1136 is $\alpha_{2000}$ = 11:39:05.570,
+66:30:17.75.  The wide companion is separated by $\mu$ = 1.349'' at
$\theta$ = 54.391$^o$, or at position 11:39:05.945, +66:30:18.45.
Since the \emph{STIS} imaging used the broad-band clear imaging 50CCD
mode, the band pass encompasses a wavelength range from 2000\AA\ to past
1$\mu$, though the sensitivity peaks around the $B$,$V$ and $R$ bands.
The relatively brightness or count ratio of HS~1136AB to the resolved
companion is 5.5:1 (nearly 2 magnitudes).

\begin{figure}

\plotone{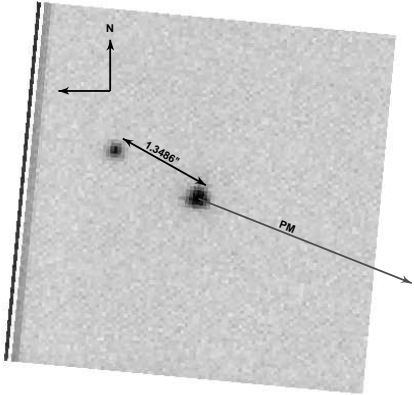}
\caption{The acquisition image with the \emph{STIS} spectrograph on the
\emph{Hubble Space Telescope}.  The red companion is fully-resolved at
the separation marked in the figure and at the orientation indicated in
the previous figures.  The \emph{STIS} wavelength band is very broad but
centered near $V$.  The companion is 1.9 magnitudes fainter than the
HS~1136 binary.  The proper motion measurement is indicated by the
arrow, and discussed in the text}

\end{figure}

The measured proper motion from the USNO-B Catalog is 3.23'' in 50 years
at a position angle of 248.2 degrees, the wide companion to the
close binary is not resolved.  If we assume that the resolved companion
is an unrelated field star with negligible proper motion, the two stars
are presently moving apart.  If we consider the POSS~I fields (epoch
1954.1), the red and blue images are symmetrical and coincident with no
evidence of an offset, even though the proper motion difference between
HS~1136 and a stationary companion would have resulted in a 1.9''
separation on the opposite side of HS~1136.  However, the effective
image size at the Palomar Ochsen Schmidt scale of 67''/mm is several
times this separation.  Thus, the companion has not been demonstrated to
share a common proper motion.

Estimating the 3.84 years of proper motion between the SDSS and STIS
ACQ images for the close binary and the resolved component, shows each
has moved around 0.25'' in roughly similar directions (close binary:
$\mu$ = 0.23'' / 3.84 yr, $\theta$ = 239$^{\circ}$; resolved
component: $\mu$ = 0.26'' / 3.84 yr, $\theta$ = 288$^{\circ}$).
Note that what we call the resolved component is close, but not
technically resolved by the Rayleigh criteria in the $z$ band image,
complicating its proper motion calculation.

We have retrieved the SDSS $u$ $g$ $r$ $i$ $z$ magnitudes corresponding
to the image of HS1136+6646 using the Explore tool in Sky Server.  The
resolved component is virtually undetected in $u$, but is nearly
resolved in the $z$ band image.  Sky server has separated the two
components into the close binary SDSS J113905.78+663017.8 and the
resolved component SDSS J113905.98+663018.3.  Point spread function
(PSF) photometry from the SDSS website lists $u$ = 13.298$\pm$ 0.014,
$r$ = 13.914$\pm$ 0.03, $i$ = 14.084$\pm$ 0.08, and $z$ = 14.159$\pm$
0.034 for the close binary and $u$ = 18 $\pm$6, r = 16.12$\pm$0.31, i =
14.457$\pm$0.048, and $z$ = 14.052$\pm$0.028 for the resolved component.
The $g$ band flux is saturated and therefore not used.  The $z$ band
flux from the close binary and resolved component can be estimated by
convolving a synthetic spectrum of the white dwarf (Sing 2005) with the
z band filter and applying the photometric corrections of Holberg \&
Bergeron (2006).  We find that the white dwarf ``A'' component has an
estimated $z$ magnitude of 15.2 while the binary companion ``B'' is
14.7.

Unfortunately, the separate PSF magnitudes other than $z$ are
unreliable.  This is illustrated by the colors of the resolved
component ( $r-i$ = 1.659 and $i-z$ = 0.432) which do not give main
sequence colors.  The $z$ magnitudes of the two main sequence stars
are observed to be similar and should therefore both contribute to the
overall flux of HS1136+6646 in the spectroscopic and photometric
observations reported by Sing et al.  (2004).  These authors showed
that the secondary star in HS1136+6646 (which we now know is a
composite of two stars) was consistent with a K7 V from the optical
down to the infrared.  The two main sequence stars, with similar $z$
magnitudes, would therefore have to be of similar K spectral type to
give these consistent colors across the wide spectral range.  

Analyzing the photometry of the 50CCD image can provide further hints as
to the spectral type of the HS~1136C component.  We convolved the clear
band pass of the 50CCD image with the observed spectral energy
distributions of (1) the white dwarf primary, (2) a K7~V close binary
with $z$=14.7, and (3) a resolved G-M component with $z$=14.025.  A
synthetic brightness ratio of HS~1136AB to HS1136C was then calculated
for different resolved companion spectral types and compared with the
observed 5.5:1 ratio.  We find a K4~V or K5~V is most consistent for
HS~1136C giving ratios of 5.3:1 and 5.8:1 respectively, while G or M
types are ruled out with a G5~V giving a ratio of 3.3:1 while a M3~V
gives 12.9:1.  Even a K3~V with a ratio of 4.7 or a K7~V with 7.0 are
unlikely matches.  With the main sequence stars having similar
magnitudes, type, and proper motion, it would seem likely that
HS1136+6646 is a triple star system, although further studies will be
needed for a definitive proof.

Note that Tokovinin et al. (2006) found via adaptive optics
observations that 13 of 62 solar-type spectroscopic binaries have
tertiary companions.  Similar observations of HS~1136AB and its
neighbor with an AO system should be sufficient to establish more
accurately the color and likely spectral type of the latter, and
establish whether its magnitude is consistent with it being a
companion at similar distance.

\section {Conclusions}

We failed to detect any diffuse nebula surrounding the young post-CE
system HS~1136.  We did, however, uncover evidence of an apparent
third component contributing to the optical light.  It is prsently
unclear if this component, easily resolved in the \emph{HST}
acquisition image, is an unrelated field star or a common proper
motion companion making a hierarchial triple system.  Evidence
indicates that it is likely of K spectral type.  Regardless,
ground-based spectroscopic and photometric studies of this systems
need to take into account the presence of the additional star.

\acknowledgments

We thank Richard Cool and Ed Olszewski for providing the H$\alpha$ flat
fields for the \emph{90Prime} images. JBH wishes to acknowledge support
from Space Telescope Science Institute grant GO09762.  JL and KAW 
acknowledge support from the National Science Foundation through grant 
AST03-07321.

\end{document}